\documentstyle[12pt,epsfig]{article}
\renewcommand{\bar}[1]{\overline{#1}}

\begin{document}

\begin{flushright}
GEF-Th-24/2006\\
\end{flushright}

\bigskip\bigskip
\begin{center}
{\large \bf The QCD Parton Model: a Useful Approximation}
\end{center}
\vspace{12pt}

\begin{center}
 {\bf Elvio Di Salvo\\}

 {Dipartimento di Fisica and I.N.F.N. - Sez. Genova, Via Dodecaneso, 33 \\-
 16146 Genova, Italy\\}
\end{center}

\vspace{10pt}
\begin{center} {\large \bf Abstract}

Approximate relations among transverse momentum dependent quark distribution functions
are established in the framework of the QCD parton model. The validity of such results
survives QCD evolution effects, owing to the Politzer theorem on equations of motion. 
Furthermore the model fixes an energy scale, involved in the parametrization of the 
correlator, which determines the $Q^2$ dependence of the azimuthal asymmetries in inclusive 
reactions. Some of the present data - the $cos2\phi$ asymmetry in unpolarized Drell-Yan, 
the $sin2\phi$ single spin asymmetry in semi-inclusive deep inelastic scattering (SIDIS) 
and the $cos \phi$ asymmetry in unpolarized SIDIS - support model predictions.  Further 
measurements of SIDIS and Drell-Yan asymmetries are suggested, in particular the double spin 
asymmetry in semi-inclusive deep inelastic scattering, which allows to determine 
approximately the proton transversity. 

\end{center}

\vspace{10pt}

\centerline{PACS numbers: 13.85.Qk, 13.88.+e}

\newpage

\section{Introduction}

The quark-quark correlator\cite{rs,mt} is a fundamental tool for calculating cross sections 
in inclusive high energy reactions. It may be parametrized according to the Dirac algebra, 
using the available vectors. We distinguish between the common correlator, which depends only 
on the longitudinal fractional momentum, and the transverse momentum dependent one. In the 
former case the correlator consists of 9 components\cite{jj}, while in the latter case the 
number of independent functions is 32\cite{bam,gms} - two for any Dirac operator - of which 
23 are washed out upon integration over transverse momentum. These last functions contain the 
dimensionless four-vector 
\begin{equation}
\eta_{\perp} = p_{\perp}/{\mu_0}, \label{etap}
\end{equation}
where $p_{\perp}$ is the transverse four-momentum of the active quark with respect to the
hadron and $\mu_0$ an undetermined energy scale, introduced for dimensional reasons. We shall 
fix this parameter by comparison with the limiting case of the QCD parton model, which 
provides also useful approximate relations between different soft functions. Some predictions 
of this model are tested against data and further measurements are suggested.

First of all we introduce the quark-quark correlator, $\Phi$, discussing its normalization. 
Secondly, we parametrize $\Phi$ according to the Dirac components. Then we write the density 
matrix of a quark in the approximation of the QCD parton model. Comparison with the general 
parametrization of $\Phi$ allows to determine $\mu_0$ and implies approximate relations among 
transverse momentum distributions. Lastly we compare our results with present data and 
suggest for them further experimental tests.

\section{Parametrization of the correlator}

The correlator is involved in a typical calculation of inclusive cross section - for example, 
deep inelastic scattering (DIS), semi-inclusive DIS (SIDIS), or Drell-Yan (DY) - {\it i. e.}, 
\begin{equation}
\frac{d\sigma}{d\Gamma} =
\frac{(4\pi\alpha)^2}{4{\cal F}Q^4} L^{\mu\nu} W_{\mu\nu}.  \label{thecs}
\end{equation}
Here $d\Gamma$ is the phase space element, $\alpha$ the fine structure constant, $Q^2$ the 
virtuality
of the exchanged photon and ${\cal 
F}$ the flux factor. Moreover $L^{\mu\nu}$ and $W^{\mu\nu}$ are respectively the leptonic and 
hadronic tensor. In particular, the hadronic tensor reads, at zero order in QCD,
\begin{equation}
W^{\mu\nu} = c\sum_ae^2_a\int d^2p_{\perp} Tr\left[\Phi^a_A(x_a, {\bf p}_{\perp}) 
\gamma^{\mu}\Phi^b_B(x_b, {\bf q}_{\perp}-{\bf p}_{\perp}) \gamma^{\nu}\right]. \label{ht}
\end{equation}
Here $c = 1$ and the flavor $b$ coincides with $a$ for DIS (and SIDIS), while $c = 1/3$ and  
$b = {\bar a}$ for DY. Each correlator is normalized so as to reduce to the usual spin 
density matrix in absence of parton-parton interactions. For a nucleon we have
\begin{equation}
\Phi = \Phi_e + \Phi_o, \label{eo}
\end{equation}
where $\Phi_{e}$ is even under time reversal and $\Phi_{o}$ is odd under the same 
transformation. $\Phi_{e}$ reads, up to and including twist 3\cite{mt,bhm,gms},
\begin{equation}
\Phi_e \simeq \Phi_e^{(2)} + \Phi_e^{(3)}, \label{ee}
\end{equation}
with 
\begin{eqnarray}
\Phi_e^{(2)} &=& \frac{\cal P}{\sqrt{2}}  \{f_1\rlap/n_+ + (\lambda 
g_{1L}+\lambda_{\perp}g_{1T})\gamma_5\rlap/n_+ + 
\frac{1}{2}h_{1T}\gamma_5[\rlap/S_{\perp},\rlap/n_+]\nonumber
\\ 
&+& \frac{1}{2}(\lambda h^{\perp}_{1L}+\lambda_{\perp} h^{\perp}_{1T}) \gamma_5 
[\rlap/\eta_{\perp},\rlap/n_+]\}\label{par02}
\end{eqnarray}
and
\begin{eqnarray}
\Phi_e^{(3)} &=&  \frac{1}{2}(f^{\perp}+\lambda g^{\perp}_L\gamma_5 + \lambda_{\perp} 
g^{\perp}_T\gamma_5)\rlap/p_{\perp} + \frac{1}{4}\lambda_{\perp} h_T^{\perp} \gamma_5 
[\rlap/S_{\perp}, \rlap/p_{\perp}] \nonumber
\\
&+& \frac{1}{2}x M \{e + g'_T\gamma_5 \rlap/S_{\perp} + \frac{1}{2}(\lambda h_L  + 
\lambda_{\perp} h_T) \gamma_5 [\rlap/n_-,\rlap/n_+]\}. \label{par03}
\end{eqnarray}
On the other hand, the twist-2 component of $\Phi_o$ amounts to\cite{bm,bjm}
\begin{equation}
\Phi_o \simeq \frac{\cal P}{\sqrt{2}}\{f^{\perp}_{1T} \epsilon_{\mu\nu\rho\sigma} 
\gamma^{\mu}n_+^{\nu}\eta_{\perp}^{\rho}S_{\perp}^{\sigma}+
ih^{\perp}_1\frac{1}{2}[\rlap/\eta_{\perp},\rlap/n_+]\}. \label{cmo}
\end{equation}
In formulae (\ref{par02}) to (\ref{cmo}) we have used the notations of refs.\cite{mt,tm} for 
the "soft" functions, but with different normalizations, as we shall specify below. They are 
functions of $x$ and of $p^2_{\perp}$, where $x = p^+/P^+$ and $p_{\perp}$ are, respectively, 
the longitudinal fractional momentum and the transverse four-momentum of the quark. We have 
chosen a frame where, in light cone coordinates, the four-momentum of the nucleon is $P\equiv 
(P^+,P^-,{\bf 0})$, the quark four-momentum is $p \equiv (p^+,p^-,{\bf p}_{\perp})$ and 
$p_{\perp} \equiv (0,0,{\bf p}_{\perp})$. $n_{\pm}$ are lightlike vectors, such that 
$n_+\cdot n_- = 1$, whose space components are directed along (+) or opposite to (-) the 
nucleon momentum. Moreover $P^2 = M^2$ and
\begin{equation}
S =  \lambda \frac{P}{M} +S_{\perp}\label{pl}
\end{equation}
is the Pauli-Lubanski (PL) vector of the nucleon; one has $S^2 = -1$, $\lambda = -S\cdot n_0$ 
and $n_0 \equiv 1/\sqrt{2}(1, -1, {\bf 0})$ in the nucleon rest frame. Thirdly, 
\begin{equation}
{\cal P} = \frac{1}{\sqrt{2}}p\cdot n_-, \ ~~~~~ \ ~~~~~ \ \lambda_{\perp} = -S\cdot 
\eta_{\perp},\label{li}
\end{equation}
$\eta_{\perp}$ being given by eq. $\ref{etap}$. Lastly, the energy scale $\mu_0$, encoded in 
$\eta_{\perp}$, has been introduced in such a way that all functions involved in the 
parametrization of $\Phi$ have the dimensions of a probability density. This scale - defined 
for the first time in ref.\cite{ko}, where it was denoted by $m_D$ - determines the 
normalization of the functions which depend on $\eta_{\perp}$; therefore $\mu_0$ has to be 
chosen in such a way that these functions may be interpreted just as probability densities. 
We can reasonably assume this parameter to be independent of the perturbative interactions 
among partons.

We conclude this section with a comparison between the normalization of our functions and the 
one of Mulders et al.. We distinguish between the functions which are not multiplied by 
$\eta_{\perp}$ and those which are multiplied by that vector. As regards the former category, 
we have, {\it e. g.}, 
\begin{equation}
f_1 = f_1^m/x, ~~~~~ \ ~~~~~ g_{1L} = g_{1L}^m/x, ~~~~~ \ ~~~~~ h_{1T} = h_{1T}^m/x, 
\end{equation}
where the superscript $m$ refers to the Mulders convention. Concerning the second group of 
functions, the relations are of the type, for example,
\begin{equation}
g_{1T} = g_{1T}^mM/(\mu_0x), ~~~~~ \ ~~~~~   h^{\perp}_{1L} = h^{\perp m}_{1L}M/(\mu_0x). 
\label{norm2}
\end{equation}

\section{Density matrix in QCD parton model}

Now we write the density matrix of a confined quark, but free of interactions with other 
partons. To this end, we inspire to the one of a free spin-1/2 particle, {i. e.}, 
\begin{equation}
\rho_f = \frac{1}{2}(\rlap/p+m)(1+\gamma_5\rlap/S'), \label{fr-dm}
\end{equation}
where $p$, $m$ and $S'$ are respectively the 4-momentum, the mass and the Pauli-Lubanski 
4-vector of the fermion. We rewrite eq. (\ref{fr-dm}) as
\begin{equation}
\rho_f = \frac{1}{2}(\rlap/p+m)(1+\gamma_5\rlap/S'_{\parallel}+\gamma_5\rlap/S'_{\perp}), 
\label{fdm1}
\end{equation}
where $S'_{\parallel} = \lambda' p/m$, $S'_{\perp} = S'-S'_{\parallel}$, 
$\lambda' = {\hat{\bf p}}\cdot{\hat{\bf s}}$ and ${\hat{\bf p}}$ and ${\hat{\bf s}}$ are unit 
vectors in the directions, respectively, of the quark momentum and of the spin in the 
particle rest system. In order to generalize eq. (\ref{fdm1}) to the case of interest, we 
take into account three elements:

a) the internal structure of the nucleon is sensitive both to longitudinal and to transverse 
polarization, in an independent way;

b) the spin of a massive particle has to be defined in its rest frame, therefore the PL 
vector of the quark does not coincide with the one of the nucleon\cite{ael};

c) the quark can be treated as if it were on shell, owing to the equations of motion; we 
shall discuss this point below.  

Then
\begin{equation}
\rho_c = \frac{1}{2}(\rlap/p+m_q)\left[q(x, p_{\perp}^2)+ \Delta q(x, p_{\perp}^2)\gamma_5 
\rlap/S^q_{\parallel}+\Delta_T q(x, p_{\perp}^2) \gamma_5 \rlap/S^q_{\perp}\right]. 
\label{dm3}
\end{equation}
Here $S^q_{\parallel}$ and $S^q_{\perp}$ are defined in such a way that, in the {\it quark} 
rest frame, they coincide respectively with $S_{\parallel} = S - S_{\perp}$ and $S_{\perp}$ 
in the nucleon rest frame. Therefore
\begin{equation}
S^q_{\parallel} = \lambda(\frac{p}{m_q}-\bar{\eta}_{\perp}) + O(\bar{\eta}^2_{\perp}),
~~~~~~~~~~~ \ ~~~~~~ S^q_{\perp} = S_{\perp} + \bar{\lambda}_{\perp}\frac{p}{m_q} + 
O(\bar{\eta}^2_{\perp}),  \label{qspin}
\end{equation}
with $\bar{\eta}_{\perp} = p_{\perp}/{\cal P}$ and $\bar{\lambda}_{\perp} = 
-S\cdot\bar{\eta}_{\perp}$.
Moreover, we have set
\begin{eqnarray}
q(x, p_{\perp}^2) &=& q_+(x, p_{\perp}^2)+q_-(x, p_{\perp}^2) = q_{\uparrow}(x, 
p_{\perp}^2)+q_{\downarrow}(x, p_{\perp}^2),
\\
\Delta q(x, p_{\perp}^2) &=& q_+(x, p_{\perp}^2)-q_-(x, p_{\perp}^2), ~~~~~~~~~~~~ \ 
~~~~~~~~~ \ ~~~~~~~~~~~ \ 
\\
\Delta_T q(x, p_{\perp}) &=& q_{\uparrow}(x, p_{\perp}^2)-q_{\downarrow}(x, p_{\perp}^2). 
~~~~~~~~~~~~ \ ~~~~~~~~~ 
\end{eqnarray}
Here $q_{\pm}(x, p_{\perp}^2)$ and $q_{\uparrow(\downarrow)}(x, p_{\perp}^2)$ are  the 
probability densities of finding a quark with, respectively, a positive (+) or negative (-) 
helicity and a positive ($\uparrow$) or negative $(\downarrow)$ transversity, the sign of the 
latter being determined by the sign of the scalar product between the quark transversity and 
the nucleon transversity. Eq. (\ref{dm3}) can be obtained as a limiting expression\cite{ds1} 
of the correlator in a gauge theory, {\it i. e.},
\begin{equation}
\Phi = \int \Phi' (p; P,S)dp^-,\label{corr}
\end{equation}
where the matrix elements of $\Phi' (p; P,S)$ are defined as
\begin{equation}
\Phi'_{ij}(p; P,S) = \int\frac{d^4x}{(2\pi)^4} e^{ipx} 
\langle P,S|\bar{\psi}_j(0) {\cal L}(x)  \psi_i(x)|P,S\rangle. \label{corr1}
\end{equation}
Here $\psi$ is the quark field and $|P,S\rangle$ is a state of the nucleon. Moreover
\begin{equation}
{\cal L}(x) = {\mathrm P} exp\left[-ig\Lambda_{\cal P}(x)\right], \ ~~~~~ \ {\mathrm with}
\ ~~~~~ \ \Lambda_{\cal P}(x) = \int_0^x \lambda_a A^a_{\mu}(z)dz^{\mu},\label{link}
\end{equation}
is the gauge link operator. Here "P" denotes the path-ordered product along the integration 
contour ${\cal P}$, $\lambda_a$ and $A^a_{\mu}$ being respectively the Gell-Mann matrices and 
the gluon fields. The link operator depends on the choice of ${\cal P}$, which has to be 
fixed so as to make a physical sense. However, for our aims we can neglect the details of the 
contour. Indeed, in the limit for $g \to 0$, that is, for noninteracting quarks, ${\cal 
L}(x)$ tends to 1. Moreover, in that limit, the quark may be treated as an on-shell particle, 
as shown also by Qiu\cite{qiu} via equations of motion\cite{pol}; see also ref.\cite{ds2}. It 
can be shown\cite{ds1} that in this limit $\Phi$ tends to $\rho_c$. Another important 
consequence of the equations of motion is that they survive renormalization, therefore eq. 
(\ref{dm3}) is a good approximation to the correlator even if QCD evolution of the 
distribution functions is taken into account. 

\section{Approximate equalities among "soft" functions}

Now we compare the density matrix (\ref{dm3}) with the general parametrization (\ref{par02}) 
of the twist-2, T-even component of $\Phi$. To this end we consider projections of both 
matrices over the various Dirac components, {\it i. e.}, for a given Dirac operator $\Gamma$,
\begin{equation}
\Phi^{\Gamma} = \frac{1}{2}Tr{\Gamma\Phi}.
\end{equation}
First of all, $\Gamma$ = $\gamma^+$, $\gamma_5\gamma^+$ and $\gamma_5\gamma^+\gamma_i$ ($i$ = 
1, 2) yield 
\begin{equation}
f_1 = q, ~~~~~ \ ~~~~ g_{1L} = \Delta q, ~~~~~ \ ~~~~ h_{1T} = \Delta_T q \label{ff0}
\end{equation}
and 
\begin{equation}
h^{\perp}_{1L} \approx -\frac{\mu_0}{\cal P}\Delta q, ~~~~~ \ ~~~~ g_{1T} \approx 
h_{1T}^{\perp}\approx \frac{\mu_0}{\cal P}\Delta_T q.\label{ff1}
\end{equation}
The last equalities hold approximately in the limit of $m_q$ = 0. In order to determine 
$\mu_0$, we observe that the functions $g_{1T}$, $h^{\perp}_{1L}$ and $h_{1T}^{\perp}$, 
involved in formulae (\ref{ff1}), are twist 2, therefore they may be interpreted as quark 
densities, as well as $\Delta q$ and $\Delta_T q$. For example, $g_{1T}$ is the helicity 
density of a quark in a tranversely polarized nucleon. Therefore it is natural to fix $\mu_0$ 
in such a way that such functions are normalized like $\Delta q$ or $\Delta_T q$. This 
implies, neglecting the quark mass,
\begin{equation}
\mu_0 = {\cal P} = \frac{1}{\sqrt{2}}p\cdot n_-. \label{mu0}
\end{equation}
This result differs from the treatments of previous authors\cite{mt,bhs}, who assume $\mu_0 = 
M$. Moreover, it is interesting to consider also the projections over twist-3 operators, in 
particular $\Gamma$ = $\gamma_i$ ($i$ = 1, 2). This yields
\begin{equation}
f^{\perp} \approx f_1 = q,\label{ff2}
\end{equation}
which is known as the Cahn effect\cite{ca,ca2}. We have neglected quark-gluon interac- 
tions\cite{mt}, which could modify, in principle, equalities (\ref{ff0}), (\ref{ff1}) and 
(\ref{ff2}), in particular the last one, concerning a twist 3 operator. However, as we shall 
see, comparison with data, wherever possible, suggests that such equalities are approximately 
verified.  On the contrary, the projection over $\Gamma$ = $\gamma_5\gamma_i$ ($i$ = 1, 2) 
yields (after integration over ${\bf p}_{\perp}$)
\begin{equation}
g_T(x) = \frac{m_q}{xM} h_1(x). \label{ff3}
\end{equation}
In this case the contribution of the QCD parton model is very small: $m_q$ is negligible for 
$u$- and $d$-quarks, while for $s$-quarks $h_1$ is predicted to be small, because sea quarks 
are produced mainly by  annihilation of gluons, whose transversity is zero in a nucleon. 
Therefore the contribution of quark-gluon interactions, neglected in our model, becomes 
prevalent in this case, as well as for $\Gamma$ = $1$ and $\gamma_5\gamma_+\gamma_-$, 
corresponding respectively to $e$ and $h_{L}$. Such interactions will be discussed below. Eq. 
(\ref{ff3}) - similar to some expressions given in the literature\cite{cpr,mt} - establishes 
a relation 
between transversity and transverse spin. Indeed, the two quantities are related to each 
other, as can be seen also from eq. (\ref{fdm1}) for a free fermion. But, unlike 
transversity, the transverse spin operator is chiral even and does not commute with the free 
hamiltonian of a quark\cite{jj}: in QCD parton model it is proportional to the quark rest 
mass, which causes chirality flip.  

\begin{figure}[htb]
\begin{center}
\epsfig{file=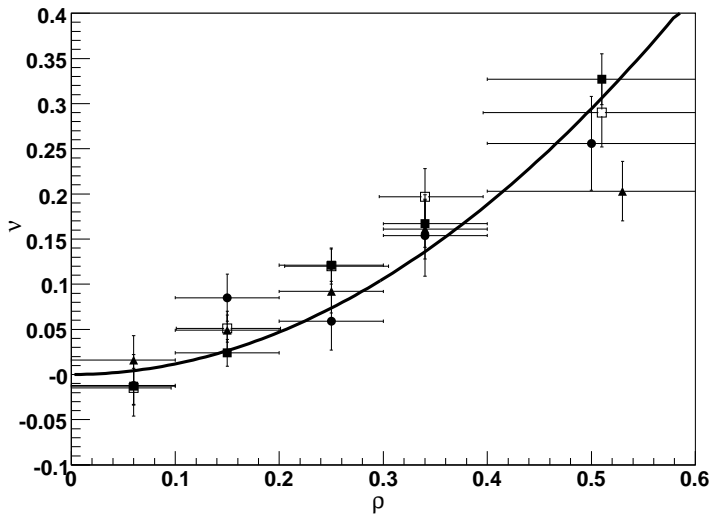,height=7.5cm}
\caption{Azimuthal asymmetry in unpolarized DY: the asymmetry parameter $\nu$ {\it vs} $\rho$ 
= $|{\bf q}_{\perp}|/Q$. Data are taken from the first two refs.\cite{fa}: circles  
correspond to $\sqrt{s}$ = 16.2 GeV, squares to $\sqrt{s}$ = 19.1 GeV and triangles to 
$\sqrt{s}$ = 23.2 GeV. The best fit is made by means of formula (\ref{dy}), taking into 
account eq.  (\ref{mu1}). $A_0$ = 1.117.} 
\end{center}
\label{fig:one}
\end{figure}

Applying eq. (\ref{mu0}) to SIDIS and DY yields
\begin{equation}
\mu_0 = Q/2. \label{mu1}
\end{equation}
Moreover, results (\ref{ff0}) to (\ref{ff2}) can be extended, with some caution\cite{bam}, to 
the fragmentation correlator.

It is worth comparing our approach to previous ones. Kotzinian\cite{ko} starts from the 
approximate expression of the density matrix for a free ultrarelativistic fermion and adapts 
it to the case of a quark in the nucleon. He parametrizes the density matrix with the 6  
twist-2, T-even functions that appear in the parametrization (\ref{dm3}). Similar results are 
obtained by Ralston and Soper\cite{rs} and by Tangerman and Mulders\cite{tm}. The difference 
with our approach is that those authors do not take into account the Politzer theorem, which 
implies relations among the 6 soft funtions. In any case, eq. (\ref{dm3}) turns out to give 
the same result as found by Ralston and Soper\cite{rs} as regards the parametrization of the 
DY hadronic tensor in the limit of parton model. Indeed, according to our approximation, the 
{\it symmetric} hadronic tensor for two polarized protons consists of just three 
coefficients, corresponding, respectively, to the convolutive products $q\otimes {\bar q}$, 
$\Delta q\otimes  \Delta{\bar q}$ and $\Delta_T q\otimes  \Delta_T{\bar q}$.   

A remark is in order. Eqs. (\ref{ff0}), (\ref{ff1}) and (\ref{ff2}) have been deduced in the 
framework of the QCD parton model. They survive QCD evolution, but are modified by 
quark-gluon interactions. In particular, the Politzer theorem implies\cite{ds2} that such 
interactions modify those approximate equalities by terms of order $g^2/Q^2$. This prediction  
could be tested, for example, in the case of $g_T$, by performing measurements at different 
$Q^2$ values. 

\section{Phenomenological analysis}

Now we test the relations just found against results from three different experiments.
 
1) Unpolarized Drell-Yan with a fixed transverse momentum ${\bf q}_{\perp}$ of the final muon 
pair. In this case the momentum of each muon presents an azimuthal asymmetry with respect to 
the plane containing the initial beam direction and ${\bf q}_{\perp}$. This azimuthal 
asymmetry is essentially of the type $1/2\nu sin^2\theta cos 2\phi$\cite{fa}, where $\theta$ 
and $\phi$ are respectively the polar and azimuthal angle of the $\mu^+$ momentum in the 
Collins-Soper frame\cite{cs}, defined in the center of mass of the pair, while $\nu$ is a 
dimensionless coefficient. This asymmetry is interpreted\cite{fa} in terms of quark-antiquark 
annihilation into a virtual photon. In the formalism of the correlator\cite{mt,bjm,ds1} the 
asymmetry is described\cite{bbh} as a convolutive product of the transversities $h_1^{\perp}$ 
and ${\bar h}_1^{\perp}$ of the two active partons in the initial hadrons, providing for 
$\nu$ an expression of the type\cite{ds1} 
\begin{equation}
\nu = A_0 \frac{{\bf q}_{\perp}^2}{\mu_0^2}. \label{dy}
\end{equation}
But eqs. (\ref{mu0}) and (\ref{mu1}) imply $\mu_0$ = $1/2 Q$, where in this case $Q$ has to 
be identified with the effective mass of the muon pair. Of course, also the first order 
perturbative QCD corrections (Compton scattering and gluon production\cite{fa}) give
rise to an asymmetry parameter which, for $|{\bf q}_{\perp}| << Q$, is again of the type 
${\bf
q}^2_{\perp}/Q^2$; however, these perturbative contributions fulfil the Lam-Tung 
relation\cite{ltu}, which, instead, results to be rather strongly violated\cite{fa}.
Figs. 1 and 2 show comparison between formula (\ref{dy}) and the parameter $\nu$, as results 
from best fits to Drell-Yan data\cite{ds1}. In this connection we observe that, owing result 
(\ref{mu1}), the $Q^2$-dependence for DY single spin asymmetry with a tranversely polarized 
nucleon coincides with the one found in ref. \cite{bmt}.

\begin{figure}[htb]
\begin{center}
\epsfig{file=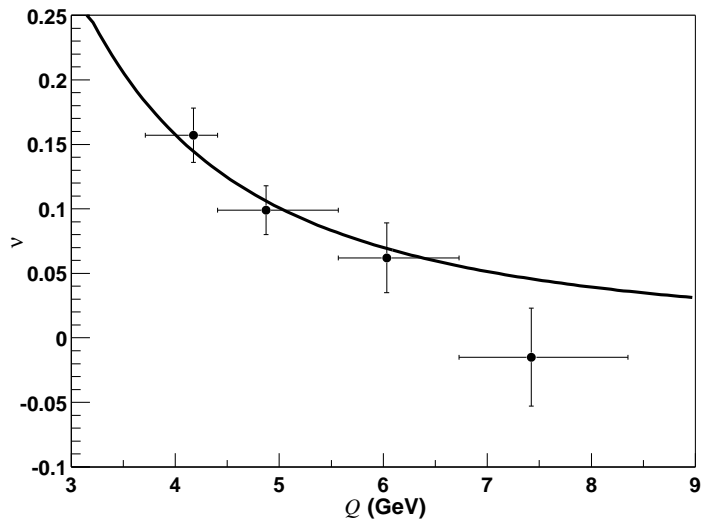,height=7.5cm}
\caption{Azimuthal asymetry in unpolarized DY: the asymmetry parameter $\nu$ {\it vs} the 
effective mass $Q$ of the final lepton pair at fixed $|{\bf q}_{\perp}|$. $\sqrt{s}$ = 23.2 
$GeV$. Data from $2^{nd}$ ref. \cite{fa} and fitted by formula (\ref{dy}), taking into 
account eq.  (\ref{mu1}). $A_0\cdot {\bf q}^2_{\perp}$ = 2.52 $GeV^2$.}
\end{center}
\label{fig:two}
\end{figure}

2) CLAS\cite{jl} results of SIDIS with longitudinally polarized beam and target. The single 
spin asymmetry includes a $sin2\phi$ term, characterized by the convolutive product 
$h_{1L}^{\perp}\otimes H_1^{\perp}$, where $H_1^{\perp}$ is the Collins\cite{co} 
fragmentation function.  The function
\begin{equation}
\tilde{h}_{1L}^{\perp}(x) = \pi \langle {\bf p}^2_{\perp}\rangle h_{1L}^{\perp}(x, {\bf 
0})\label{ltr}
\end{equation}
has been extracted\cite{jl} for {\it u}-quarks from that asymmetry, taking into account a 
model evaluation\cite{ef} of the Collins fragmentation function and assuming a Gaussian 
behavior as regards the ${\bf p}_{\perp}^2$ dependence of $h_{1L}^{\perp}(x, {\bf 
p}_{\perp}^2)$. However the normalization  adopted in ref. \cite{jl} for the "soft" functions 
is the usual one\cite{mt}, which, as already explained in sect. 2, is different from the one 
used in the parametrization (\ref{par02})-(\ref{cmo}). In particular, as regards 
$h_{1L}^{\perp}$, the conversion factor is ${\cal C} = 2M/Qx$, according to eqs. 
(\ref{norm2}) and (\ref{mu1}): it has to be taken into account, since eqs. (\ref{ff1}) hold 
within our normalization. With this re-normalization in mind, we compare (fig. 3) the 
CLAS\cite{jl} results of $-xh_{1L}^{\perp u}(x)$ [from now on, the "tilde" will be dropped 
from eq. (\ref{ltr})] to the HERMES data\cite{he1} of $xg_1^u(x)$, assuming, again, a 
Gaussian behavior for $g_{1L} (x, {\bf p}^2_{\perp})$.  The curve corresponds to the 
parametrization of $xg_1^u(x)$ given by GRSV\cite{grsv} - LO, valence scenario - scaled with 
a factor $(1+R)^{-1}$\cite{he1}. The discrepancy, which appears for $x > 0.3$, is typical of 
higher twist contribution, not negligible for the modest $Q^2$-values involved, $1.5$ to $3$ 
$(GeV/c)^2$.  

3) Unpolarized SIDIS $cos\phi$ asymmetry, $\phi$ being the azimuthal angle of the final 
hadron momentum with respect to the reaction plane. Recently Anselmino et al.\cite{ca2} 
fitted EMC\cite{emc} and E665\cite{e665} data by means of relation (\ref{ff2}). They 
parametrized $f_1(x, {\bf p}^2_{\perp})$ by making the Gaussian assumption, with $\langle{\bf 
p}^2_{\perp}\rangle$ = 0.25 $(GeV/c)^2$, and taking the MRST 2001 (LO)\cite{mrst} 
parametrization for $f_1(x)$. The agreement is good. It is to be noticed, however, that the 
values of $x$ involved are considerably smaller than those corresponding to CLAS data, 
whereas the $Q^2$-values are much greater on the average. 

\begin{figure}[htb]
\begin{center}
\epsfig{file=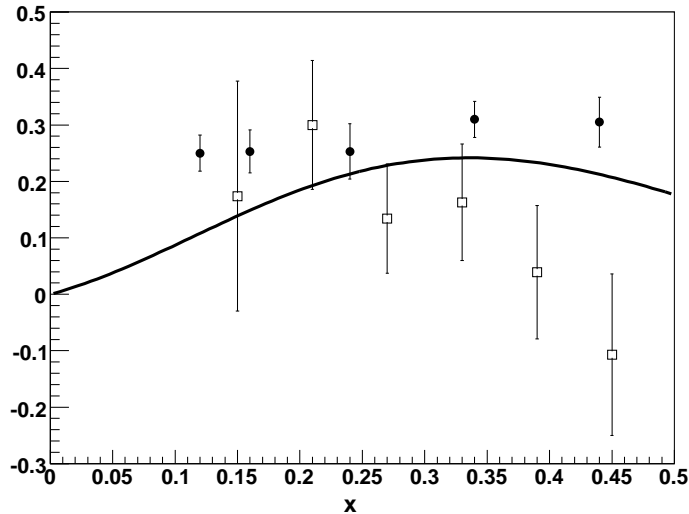,height=7.5cm}
\caption{The functions $xg_1^u(x)$\cite{he1} (full circles, $Q^2$ = 2.5 $GeV^2$) and 
$-xh_{1L}^{\perp u} $\cite{jl} (open squares, $Q^2$ = 1.5 - 3 $GeV^2$). The latter has been 
re-normalized according to the conversion factor ${\cal C} = 2M/Q$. The curve corresponds to 
the GRSV2000 parametrization (LO, valence scenario)\cite{grsv} of $xg_1^u(x)$, scaled with a 
factor $(1+R)^{-1}$\cite{he1}.}
\end{center}
\label{fig:three}
\end{figure}

\section{Conclusion and outlook}

To conclude, relations (\ref{ff0}) to (\ref{ff2}) have some predictive power, supported by 
present data. This is a goad to further measurements and investigations in forthcoming SIDIS 
and DY experiments. 
 
- First of all, we suggest to determine the $sin2\phi$ SIDIS azimuthal asymmetry with a 
longitudinally polarized target, at higher energies (12 $GeV$) and at a higher precision than 
in ref.\cite{jl}. In particular, new extractions of $h_{1L}^{\perp}(x)$ should take into 
account the recent determination of the Collins function by BELLE\cite{be} collaboration. 
According to our result, this asymmetry is predicted to decrease like $Q^{-2}$. Moreover we 
could test our prediction about higher twist effects, according to which the sum 
$h_{1L}^{\perp}(x)+g_1(x)$ decreases again like $Q^{-2}$. The function $h_{1L}^{\perp}(x)$ 
could be determined also by means of the DY single spin asymmetry, which includes a term of 
the type $h_{1L}^{\perp}\otimes {\bar h_1^{\perp}}$; in this case it would be suitable to 
realize $p{\bar p}$ collisions\cite{gs,ssn} or $\pi^-p$ scattering\cite{ss1}, since the 
function ${\bar h_1^{\perp}}$ can be extracted from unpolarized DY.

- The above determinations, while interesting in themselves, would increase reliability of 
the approximate relations deduced above, in particular, of the last two eqs. (\ref{ff1}), not 
directly testable at the moment, which can be exploited for determining approximately 
$h_{1T}$ and 
$h_{1T}^{\perp}$ through the chiral-even function $g_{1T}$. Indeed, this may be extracted 
from double spin asymmetry in SIDIS, with a longitudinally polarized beam and a transversely 
polarized target\cite{km}. No particular problems should arise in inferring $g_{1T}$ from 
asymmetry data, since this asymmetry is characterized by the convolutive product 
$g_{1T}\otimes D$, where $D$ is the common unpolarized fragmentation function of the pion.

\end{document}